\documentclass[aps,prx,twocolumn,superscriptaddress,showpacs]{revtex4-2}
\usepackage{amsmath,amssymb,wasysym,graphicx}
\usepackage[utf8]{inputenc}
\usepackage[T1]{fontenc}
\usepackage{xcolor}

\IfFileExists{newtxtext.sty}
{\usepackage{newtxtext,newtxmath}}
{\IfFileExists{stix.sty}
	{\usepackage{stix}}
	{\IfFileExists{mathptmx.sty}
		{\usepackage{mathptmx}}{} } }

\usepackage{textcomp}

\usepackage{bm}

\IfFileExists{siunitx.sty}{\usepackage{booktabs,siunitx}}{}

\usepackage{color}
\usepackage{hyperref}
\hypersetup{
	pdfauthor={good guys},
	pdftitle={good title},
	colorlinks=true,
	citecolor=blue,
	linkcolor=magenta,
	urlcolor=blue
}

\newcommand{\beq} {\begin{equation}}
\newcommand{\eeq} {\end{equation}}
\newcommand{\bea} {\begin{eqnarray}}
\newcommand{\eea} {\end{eqnarray}}
\newcommand{\be} {\begin{equation}}
\newcommand{\ee} {\end{equation}}

\usepackage{latexsym}
\usepackage{amssymb}
\usepackage{amsmath}
\usepackage{amsfonts}
\usepackage{wasysym}
\usepackage[vcentermath]{youngtab}
\usepackage{upgreek}
\usepackage{bm,dsfont}
\usepackage{multirow}
\usepackage{enumitem}
\usepackage{makecell}
\usepackage{framed}
\usepackage{color}
\usepackage{hyperref}
\usepackage{tikz}

\def\be{\begin{equation}}
\def\ee{\end{equation}}
\newcommand{\beqn}{\begin{eqnarray}}
\newcommand{\eeqn}{\end{eqnarray}}

\begin{document}

\title{Topological Superradiant Universality Classes in Cavity-Mediated Spin Chains}

\author{Yunhao Liu}
\thanks{These authors contributed equally.}
\affiliation{Eastern Institute of Technology, Ningbo 315200, China}

\author{Sheng Yang}
\thanks{These authors contributed equally.}
\affiliation{Institute for Advanced Study in Physics and School of Physics, Zhejiang University, Hangzhou 310058, China}

\author{Mingyuan Chen}
\thanks{These authors contributed equally.}
\affiliation{Eastern Institute of Technology, Ningbo 315200, China}

\author{Xue-Jia Yu}
\email{xuejiayu@eitech.edu.cn}
\affiliation{Eastern Institute of Technology, Ningbo 315200, China}

\begin{abstract} \bfseries \boldmath

Superradiant phase transitions are fundamental phenomena in light-matter coupled systems.
Here, we uncover a novel class of superradiant criticality that is topologically distinct from previously known superradiant transitions.
Using cavity-mediated cluster-Ising chains as an illustrative example, we numerically demonstrate that the critical line separating the superradiant phase from the cavity topological phase hosts robust symmetry-protected edge degeneracies, a feature absent in conventional superradiant transitions.
Remarkably, although its local bulk critical properties are indistinguishable from those of the previously identified topologically trivial superradiant transition, the two criticalities cannot be smoothly connected without breaking the protecting symmetries or passing through a multicritical point.
We further show that such topological superradiant criticality is generic, also occurring in cavity-mediated $\mathbb{Z}_{n}^{}$ cluster-qudit chains, as supported by complementary lattice simulations.

\end{abstract}

\maketitle

\noindent
The interplay between light and matter provides a powerful route to realizing collective quantum phenomena in many-body systems.
A paradigmatic example is the superradiant phase transition~\cite{HEPP1973360,Wang1973PRA,Keeling_2007,Peng2019PRA,Ye2025PRA,larson2017some,PhysRevLett.112.173601,PhysRevLett.131.113601,PhysRevA.98.063815}, in which a macroscopic ensemble of quantum degrees of freedom couples collectively to a quantized cavity mode.
As the light-matter coupling is increased, the system can undergo a transition from a normal phase with negligible photon occupation to a superradiant phase with macroscopic photon occupation and spontaneously broken parity symmetry.
These superradiant transitions are particularly intriguing because they provide a unique setting to explore how cavity-mediated interactions can reshape critical behaviors and universality classes~\cite{Tomadin10,Gammelmark2012PRA,Fan2018PRA,Guerci2020PRL,Rohn2020PRR,Mendez2020PRR,Chiocchetta2021NC,chanda2021self,Felice2022Science,Dmytruk2022CP,Chanda2022PRB,Chiriac2022PRB,Passetti2023PRL,Bacciconi2023SciPost,Weber2023CP,Zhang24,Bacciconi2024PRB,Shaffer2024PRB,Enkner2025Nature,Langheld2025PRBL,Mendonifmmode2025PRL,rao2025unilateralcriticalityphasetransition,Wei2025PRL,inácio2026motttransitionphotonsquantum,Garwola2026PRRL,Otake2026PRR,lotkov2026coupledmathbbz3rabimodels,koziol2026quantumcriticalityferromagneticdickeising,Sur2026NC,wang2026superradiantstronglycorrelatedquantum}.
Experimentally, superradiant phase transitions have been realized in a wide range of quantum platforms, including ultracold atoms and Bose--Einstein condensates coupled to optical cavities~\cite{Baumann2010Nature,Baumann2011PRL}, superconducting quantum processors~\cite{Zheng2023PRL}, and more recently solid-state magnonic systems~\cite{Dasom2025SA}.

On a different front, classifying the universal properties of quantum critical points (QCPs) has long been a central goal of condensed-matter and statistical physics.
Conventionally, continuous phase transitions are uniquely classified by a set of critical exponents, forming a standard paradigm for investigating phase transitions~\cite{landau2013statistical,Sachdev_1999,cardy1996scaling,Sondhi1997RMP}.
This paradigm has recently been challenged by the discovery of topological physics in QCPs~\cite{YU20261}.
Specifically, symmetry-protected topological edge modes, traditionally thought to exist only in gapped systems, have been shown to coexist with a gapless bulk, giving rise to the notion of critical topology~\cite{Verresen2018PRL,verresen2020topologyedgestatessurvive,Verresen2021PRX,Duque2021PRB,Yu2022PRL,Mondal2023PRB,Yu2024PRB,Prembabu2024PRB,Zhong2025PRB,Huang2025SciPost,Li2025PRB,Rey2025PRB,Zhou2025CP,chou2025ptsymmetryenrichednonunitarycriticality,guo2026generalizedlihaldanecorrespondencecritical,deng2026anomalousdynamicalscalingtopological,prembabu2025noninvertibleinterfacessymmetryenrichedcritical,yang2026topologicalquantumcriticalityquasiperiodic,TanCP2026,yang2026topologicaltricriticalisinguniversality,chou2026topologicallyenforcedlifshitzmulticriticality} or, more generally, gapless symmetry-protected topological (SPT) states~\cite{Keselman2015PRB,Scaffidi2017PRX,JIANG2018753,Parker2018PRB,Thorngren2021PRB,Hidaka2022PRB,Wen2023PRB,Yu2024PRL,Zhang2024PRA,Su2024PRB,Li2025SciPost,Wen2025PRB,Yang2025CP,Flores2025PRL,yang2025deconfinedcriticalityintrinsicallygapless,prembabu2025multicriticalitypurelygaplessspt,Yu2026PRB,xu2026frameworkpredictingentanglementspectra,chou2026topologicallyenforcedlifshitzmulticriticality,chen2026criticaltopologicalphotonicssynthetic}.
Consequently, distinct universality classes can share the same local bulk critical exponents but remain topologically distinct, and they cannot be smoothly connected without encountering a multicritical point or breaking the protecting symmetry.
This reveals a broader landscape of quantum criticality beyond the existing paradigm, where the interplay between topology and criticality gives rise to novel phenomena absent in conventional gapped states, including symmetry-enriched conformal boundary conditions~\cite{Yu2022PRL,Parker2018PRB}, unconventional bulk-boundary correspondence~\cite{Yu2024PRL,Zhong2025PRB,guo2026generalizedlihaldanecorrespondencecritical}, algebraically localized edge modes~\cite{Verresen2021PRX,Yang2025CP}, anomalous dynamical scaling~\cite{deng2026anomalousdynamicalscalingtopological}, and intrinsically gapless topological states~\cite{Thorngren2021PRB,Zhang2024PRA}.
Beyond these theoretical advances, the recent observation of critical topological phenomena in phononic metamaterials~\cite{cheng2026observe,lin2026experimental} and superconducting quantum processors~\cite{TanCP2026} has further stimulated interest in this emerging field. 

These advances naturally motivate the fundamental and intriguing question: 
can superradiant transitions be classified beyond parity symmetry breaking, and can they give rise to universality classes that are distinct from those of conventional light-matter coupled systems?
Resolving this question would be of interest not only for the theoretical development of quantum optics and condensed-matter physics, but also for experiments on modern quantum platforms.

In this work, we address this question by identifying a topologically nontrivial superradiant criticality in light-matter coupled systems.
As an illustrative example, we first consider a cavity-mediated cluster-Ising chain and numerically map out its global phase diagram, which hosts a superradiant critical line separating a superradiant phase from a cavity SPT phase protected by the joint $\mathbb{Z}_{2}^{}\times\mathbb{Z}_{2}^{T}$ symmetry of the coupled light-matter system.
More importantly, we unambiguously demonstrate that this superradiant critical line is topologically nontrivial, as manifested by robust symmetry-protected edge degeneracies.
Despite sharing indistinguishable local bulk critical properties with the previously identified topologically trivial superradiant transition, this topologically nontrivial critical line cannot be smoothly connected to its trivial counterpart without either breaking the protecting symmetries or passing through a multicritical point.
We further show that such topological superradiant criticality is not restricted to the $\mathbb{Z}_{2}^{}$ Ising case but extends to cavity-mediated $\mathbb{Z}_{n}^{}$ cluster-qudit chains.

\vspace{.5cm}
\noindent\textbf{\large{}Cavity-mediated cluster-Ising Model}

\begin{figure}[t]
  \includegraphics[width=\linewidth]{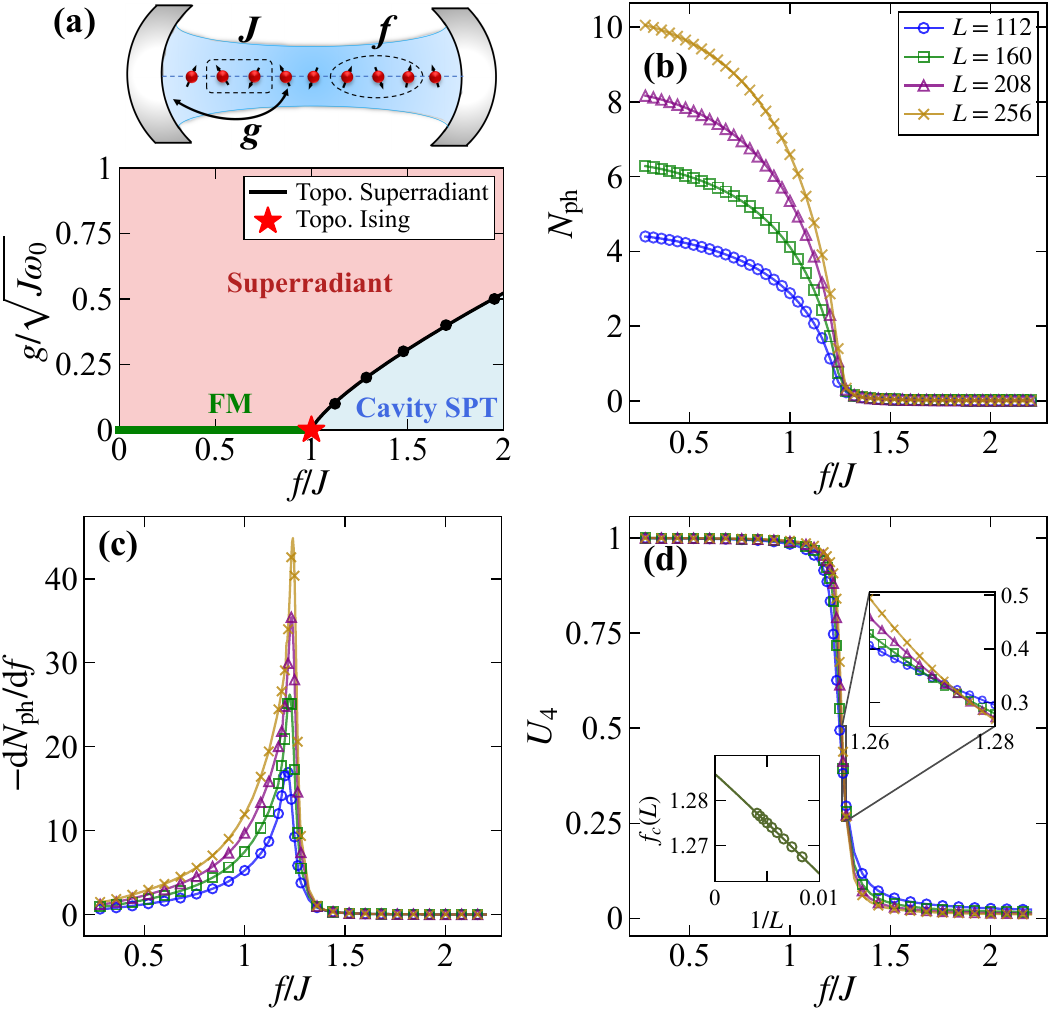}
  \caption{(a) Schematic of a cavity mode coupled to a cluster-Ising spin chain and the corresponding global ground-state phase diagram, featuring FM, superradiant, and cavity SPT phases. 
  The red star and black line denote the topological Ising critical point and topological superradiant critical line, respectively. 
  For $g/\sqrt{J\omega_0} = 0.2$, (b) photon occupation number $N_{\rm ph} = \langle a_{}^{\dagger} a_{}^{} \rangle$ and (c) its derivative with respect to $f$ as functions of $f$ for various system sizes. 
  (d) Binder cumulant of the $x$-component of the spin magnetization as a function of $f$ for various system sizes. 
  The crossing of the curves locates the corresponding finite-size critical points $f_{c}(L)$. The inset shows the scaling of the crossing points for system sizes $L=112,128,144,160,176,192,208,224,240$. 
  By extrapolating to $L \to \infty$, the thermodynamic-limit value is $f_{c}(\infty) \approx 1.286$\,.}
  \label{fig1}
\end{figure}

\noindent
We consider a spin-$1/2$ cluster-Ising chain of $L$ sites coupled to a single photon mode, described by the field operators $a_{}^{}$ and $a_{}^{\dagger}$ with frequency $\omega_{0}$.
The cavity photons couple to the $x$-component of the spin magnetization with a collective coupling constant $g$ [see Fig.~\ref{fig1}(a) for illustration].
The resulting light-qubit Hamiltonian is therefore given by
\begin{equation}
  \label{E1}
  \begin{split}
  H_{\mathrm{CI}}^{\mathrm{cavity}} & = \omega_{0} a_{}^{\dagger} a_{}^{} + \frac{g}{\sqrt{L}} \left( a_{}^{} + a_{}^{\dagger} \right) \sum_{i} \sigma^{x}_{i} + H_{\mathrm{CI}} \, , \\
  H_{\mathrm{CI}}^{} & = - J \sum_{i} \sigma^{x}_{i} \sigma^{x}_{i+1} - f \sum_{i} \sigma^{x}_{i} \sigma^{z}_{i+1} \sigma^{x}_{i+2} \, ,
  \end{split}
\end{equation}
where $\sigma_{i}^{x,z}$ are Pauli matrices at site $i$, and $J$ ($f$) denotes the Ising (cluster) interaction strength.
Without loss of generality, we set $J$ and $\omega_0$ to unity.
The model respects a spinless time-reversal symmetry $\mathbb{Z}_{2}^{T}$ generated by $T=K$ the complex conjugation, as well as a joint global $\mathbb{Z}_{2}$ symmetry generated by $\mathcal{P} = \text{e}^{\text{i}\pi a_{}^\dagger a_{}^{}} \prod_{i} \sigma_{i}^{z}$, under which $a_{}^{} \rightarrow - a_{}^{}$, $\sigma_{}^{x} \rightarrow - \sigma_{}^{x}$, $\sigma_{}^{y} \rightarrow - \sigma_{}^{y}$, and $\sigma_{}^{z} \rightarrow \sigma_{}^{z}$.
We note that the magnetic component of the light field couples to the localized spins via the Zeeman interaction, ensuring gauge invariance of the model~\cite{Kirton2019AQT,Frisk2019NRP,Razifmmode1975PRL,Nataf2010NC,Dasom2025SA} [see Supplementary Note~2 for details]. 

These nonperturbative light-qubit interactions place the model beyond standard perturbative or mean-field approaches~\cite{Peng2019PRA,Zhu2020PRL}, rendering it analytically intractable. 
We therefore employ finite-size density-matrix renormalization group (DMRG) simulations~\cite{SCHOLLWOCK201196,Verstraete2023NRP} to study the light-qubit coupled system with a truncated photon Hilbert space. 
Brief descriptions of the numerical implementation and convergence tests are provided, respectively, in Methods and Supplementary Note~1.

\vspace{.5cm}
\noindent\textbf{\large{}Global phase diagram and superradiant phase transition}

\noindent
Before presenting the detailed analysis of our DMRG results, we first summarize the ground-state phase diagram of the model in Eq.~\eqref{E1}, shown in the $(g,f)$ plane in Fig.~\ref{fig1}(a). 
At $g = 0$, the cavity and spin degrees of freedom are completely decoupled, and the spin sector reduces to a purely one-dimensional cluster-Ising chain, which undergoes a continuous transition between the ferromagnetic (FM) phase and the $\mathbb{Z}_{2} \times \mathbb{Z}_{2}$ cluster SPT phase.
The corresponding Ising critical point between them features twofold degenerate edge modes and is thus topologically nontrivial~\cite{Verresen2021PRX,Yu2022PRL,YU20261}
(see Methods for a brief introduction). 
In contrast, at $f = 0$, an infinitesimal light-spin coupling drives the system from the FM phase into the superradiant phase through a first-order transition~\cite{Sur2026NC}, with the two phases breaking fundamentally different $\mathbb{Z}_{2}^{}$ symmetries: 
the global spin-flip symmetry in the FM phase and the parity symmetry in the superradiant phase. 

In the cluster SPT region ($f/J > 1$), turning on a small light‑qubit coupling $g < g_{c}$ preserves the SPT order. 
However, the SPT phase is instead protected by time‑reversal and the joint $\mathbb{Z}_{2}$ symmetry $\mathcal{P}$ that is unique to light‑qubit systems; 
we refer to this phase as the cavity SPT phase. 
As $g$ is further increased, the ground state spontaneously breaks the $\mathbb{Z}_{2}$ parity symmetry $a \to -a$ and enters the superradiant phase, where cavity photons condense. 

In the following, we select $g/\sqrt{J\omega_0}= 0.2$ and compute the photon occupation number as a function of $f$ for various system sizes; see Fig.~\ref{fig1}(b). 
The data show that the photon occupation number vanishes continuously when crossing from the superradiant region ($f < f_{c}$) to the cavity SPT region ($f > f_{c}$). 
This superradiant transition is also clearly visible from the peaks of the derivative ($-\partial\langle a_{}^{\dagger} a_{}^{} \rangle / \partial f$) plotted against $f$ [Fig.~\ref{fig1}(c)]; 
the peak positions approach the critical point as the system size increases. 
Complementarily, since photon condensation breaks the $\mathcal{P}$ symmetry, which is also associated with the magnetic order, the precise location of the critical point can also be determined from the spin Binder cumulant $U_{4} = \frac{3}{2} \left( 1 - \frac{\langle m_{x}^{4} \rangle}{3 \langle m_{x}^{2} \rangle^{2}} \right)$~\cite{sandvik2010AIP}, with $m_{x} = \frac{1}{L} \sum_{i} \sigma_{i}^{x}$. 
As shown in Fig.~\ref{fig1}(d), $U_{4}$ exhibits a clear crossing at $f_{c}$, approaching $1$ with increasing system size for $f < f_{c}$ and $0$ for $f > f_{c}$, providing further evidence for a continuous transition associated with magnetic ordering. 
The inset shows the finite-size scaling of the crossing points obtained from system sizes $L$ and $L+16$. These finite-size pseudocritical points are extrapolated to the thermodynamic limit to estimate the critical point, giving rise to $f_{c}(\infty) \approx 1.286$\,.

Thus, the topologically nontrivial cluster-Ising critical point at $g=0$ continuously evolves into a superradiant critical line upon coupling the spin chain to the cavity. 
The central question addressed in this work is whether its topological character survives the light-qubit coupling, giving rise to a topologically nontrivial superradiant criticality.

\vspace{.5cm}
\noindent\textbf{\large{}Topological signatures at superradiant criticality}

\noindent
Recent advances~\cite{YU20261,Verresen2021PRX,Yu2022PRL} have revealed that quantum phase transitions can host nontrivial topological edge modes, giving rise to a novel form of topological phenomena known as critical topology. 
This implies that QCPs can be distinguished by their topological properties even when they share the same local bulk critical exponents.
A natural question therefore arises: is the superradiant transition identified here topologically distinct from the transition of the cavity transverse-field Ising chain reported in the literature~\cite{Sur2026NC,Rohn2020PRR,Langheld2025PRBL}?

\begin{figure}[t]
  \includegraphics[width=1.0\linewidth]{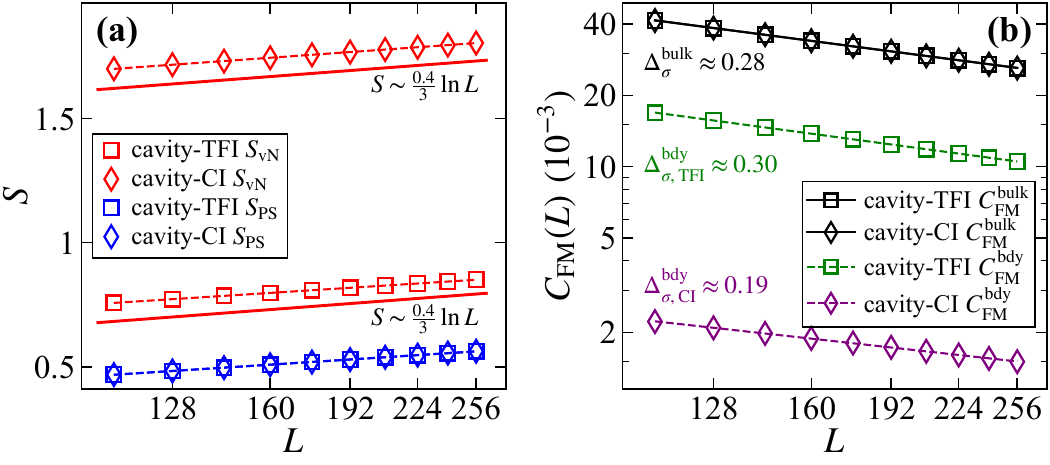}
  \caption{(a) Finite-size scaling of the entanglement entropies of the spin ($S_{\rm vN}$, red) and photon-spin ($S_{\rm PS}$, blue) sectors at the cavity transverse-field Ising (cavity-TFI, squares) and the cavity cluster-Ising (cavity-CI, diamonds) superradiant QCPs for $g/\sqrt{J\omega_0} = 0.2$\,. 
  The two red solid lines are guides to the eye for $c_{\rm eff} = 0.4$\,. 
  (b) The largest-distance ($L/2$) bulk-bulk and bulk-boundary connected spin correlations as functions of the system size $L$ at the two QCPs shown in (a), plotted on a log-log scale. Solid and dashed lines denote power-law fits.}
  \label{fig2}
\end{figure}

\begin{figure}[t]
  \includegraphics[width=1.0\linewidth]{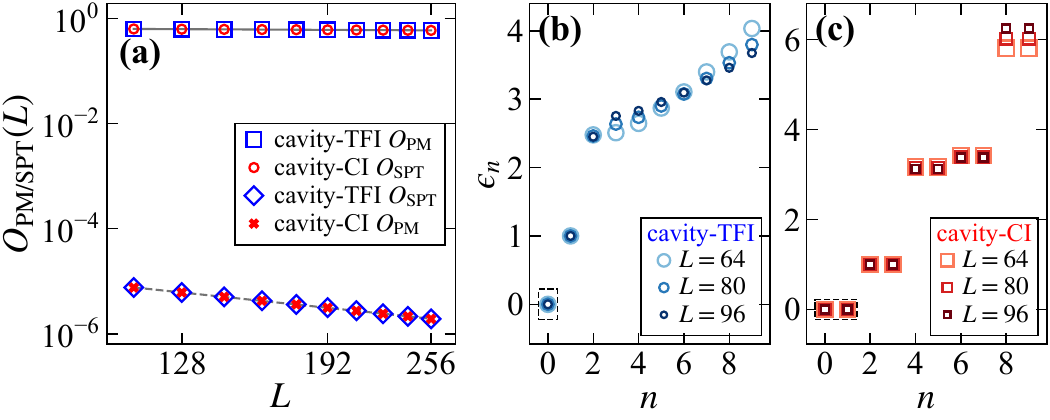}
  \caption{(a) Largest-distance ($L/2$) SPT and PM string order parameters as functions of the system size $L$ at the cavity transverse-field Ising (cavity-TFI) and the cavity cluster-Ising (cavity-CI) superradiant QCP, respectively, for $g/\sqrt{J\omega_0} = 0.2$, plotted on a log-log scale. 
  Solid and dashed lines denote power-law fits. 
  Many-body energy spectra $\{\epsilon_n\}$ for $L  = 64,80,96$ under OBC at the cavity-TFI (b) and cavity-CI (c) superradiant QCPs, with $g/\sqrt{J\omega_0} =0.2$\,.
  The energy spectra are rescaled as $\epsilon_{n} = (E_{n} - E_{0}) / (E_{1} - E_{0})$ and $\epsilon_{n} = (E_{n} - E_{0}) / (E_{2} - E_{0})$, respectively, in (b) and (c).}
  \label{fig3}
\end{figure}

A key hallmark of critical topology is that the local bulk critical properties are indistinguishable from those of a topologically trivial critical point, while the nontrivial topological features manifest in nonlocal bulk diagnostics or boundary properties. 
To demonstrate this distinction in our setting, we compare the bulk and boundary properties of the cavity cluster-Ising and cavity transverse-field Ising chains at their respective superradiant transition points.

We first examine their entanglement properties through the finite-size scaling of two complementary measures: the spatial spin entropy ($S_{\mathrm{vN}}$, red) and the photon-spin entropy ($S_{\mathrm{PS}}$, blue), with explicit definitions given in Methods.
The latter quantifies entanglement between the cavity mode and the entire spin chain.
At both transitions, $S_{\mathrm{vN}}$ grows logarithmically with system size, with closely matching prefactors [Fig.~\ref{fig2}(a)], as in conventional critical one-dimensional spin chains.
By analogy with the standard $(1+1)$-dimensional conformal-field-theory (CFT) scaling form~\cite{Pasquale_Calabrese_2004,Calabrese_2009}, we characterize the logarithmic prefactor by an effective central charge $c_{\mathrm{eff}}$.
For the conventional critical Ising chain, $c_{\mathrm{eff}} = 1/2$ coincides with the central charge of the Ising CFT, whereas here we obtain $c_{\mathrm{eff}} \approx 0.4$ at both superradiant transitions.
Interestingly, $S_{\mathrm{PS}}$ also grows logarithmically with system size, reminiscent of the critical light-matter entanglement in the Dicke model, where the logarithmic divergence is associated with collective-mode softening~\cite{Lambert2004PRL}.

The bulk features of the superradiant transitions are further corroborated by examining the bulk-bulk spin correlation function $C_{\mathrm{FM}}(L/2)$ under periodic boundary conditions (PBCs) (see Methods for definitions), as shown in Fig.~\ref{fig2}(b). 
At both critical points, $C_{\mathrm{FM}}(L/2)$ decays algebraically with system size as $C_{\mathrm{FM}}(L/2)\sim L^{-2\Delta_{\sigma}^{\mathrm{bulk}}}$, yielding the same bulk scaling dimension $\Delta_{\sigma}^{\mathrm{bulk}} \approx 0.28$, which differs from the standard Ising-CFT value $\Delta_{\sigma}^\text{bulk} = 1/8$ (see Supplementary Note~2 for the results of the bulk-bulk energy correlation). 
In contrast, the boundary critical behaviors of the two superradiant transitions are clearly distinguished by the boundary-bulk spin correlation function, evaluated under open boundary conditions (OBCs) between a boundary spin and a bulk spin located a distance $L/2$ away.
At both superradiant QCPs, the boundary-bulk spin correlation function decays algebraically as $L^{-(\Delta_{\sigma}^{\mathrm{bulk}}+\Delta_{\sigma}^{\mathrm{bdy}})}$, but the extracted boundary scaling dimensions differ markedly: 
$\Delta_{\sigma}^{\mathrm{bdy}} \approx 0.30$ for the cavity transverse-field Ising QCP and $\Delta_{\sigma}^{\mathrm{bdy}} \approx 0.19$ for the cavity cluster-Ising QCP (see Supplementary Note~2 for the results of the boundary-bulk energy correlation).
Thus, while the cavity cluster-Ising superradiant QCP identified here shares the same local bulk properties as the cavity transverse-field Ising superradiant QCP proposed in Ref.~\cite{Sur2026NC}, it exhibits fundamentally distinct boundary critical behavior. 
As discussed below, this distinction originates from the nontrivial topological edge degeneracy of the cavity cluster-Ising superradiant transition. 

\begin{figure}[t]
  \includegraphics[width=1.0\linewidth]{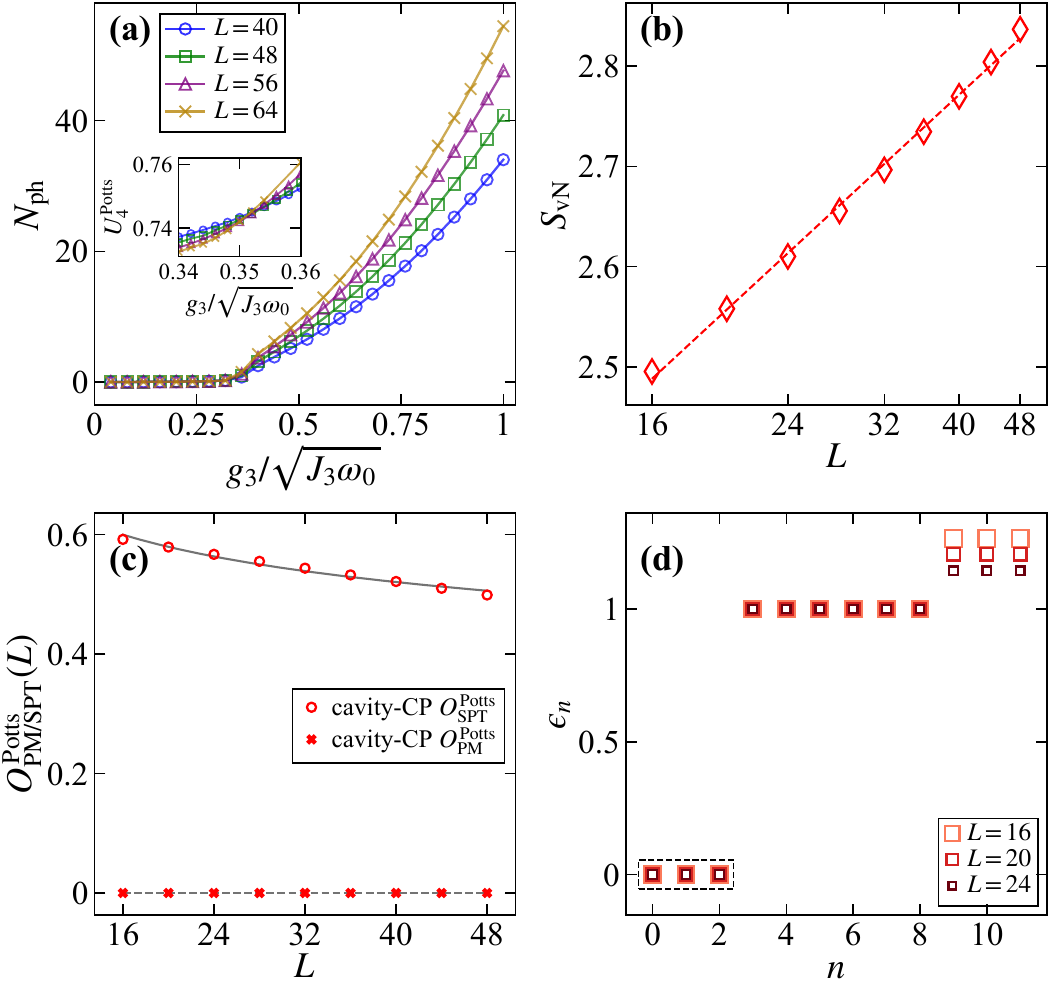}
  \caption{For fixed $f_3/J_3=1.2$: (a) Photon occupation number as a function of $g_3$ for various system sizes, showing the cavity cluster-Potts superradiant transition upon tuning $g_3$. 
  The inset shows the crossing of the qutrit Binder cumulant, which determines the critical point at $g_3/\sqrt{J_3\omega_0} \approx 0.353$\,. 
  At the corresponding QCP, (b) finite-size scaling of the entanglement entropy $S_{\rm vN}$, (c) largest-distance ($L/2$) $\mathbb{Z}_3$ SPT and PM string operators as functions of the system size $L$, and (d) the OBC many-body energy spectrum for various system sizes $L$. 
  Here, $L$ denotes the number of unit cells throughout the figure. 
  The energy levels are rescaled as $\epsilon_{n} = (E_{n} - E_{0}) / (E_{3} - E_{0})$ in (d).}
  \label{fig4}
\end{figure}

To further elucidate the topological distinction between the two superradiant transitions, we note that both critical points admit nonlocal disorder operators (denoted $\mu$) carrying distinct time-reversal symmetry charges, $T\mu T  = \pm \mu$. 
On the lattice, these operators are constructed by joint global $\mathbb{Z}_2$ symmetry $\mathcal{P}$ with different endpoint operators, corresponding to those of the pure spin chains~\cite{Verresen2021PRX}: 
$O_{\mathrm{PM}}$ for the cavity transverse-field Ising superradiant transition and $O_{\mathrm{SPT}}$ for the cavity cluster-Ising superradiant transition (see Methods for explicit expressions). 
The time-reversal charge of the $\mu$ operator thus provides a bulk topological invariant at criticality~\cite{Verresen2021PRX}, distinguishing the two superradiant transitions in the light-qubit coupled systems. 
We test this prediction numerically by evaluating the corresponding string operators, $O_{\mathrm{PM}}(L/2)$ and $O_{\mathrm{SPT}}(L/2)$, at both transitions [Fig.~\ref{fig3}(a)]. 
In the cavity cluster-Ising case, the $\mathbb{Z}_2^T$-charged string operator $O_{\mathrm{SPT}}(L/2)$ exhibits a slower algebraic decay than the $\mathbb{Z}_2^T$-neutral operator $O_{\mathrm{PM}}(L/2)$, whereas the opposite behavior is found in the cavity transverse-field Ising case. 
This distinct scaling behavior demonstrates that the two superradiant transitions are topologically distinct, reflecting qualitatively different condensation behaviors of the nonlocal $\mu$ operators. 
At the cavity Ising superradiant QCP, the condensation of the $\mathbb{Z}_2^T$-neutral $\mu$ operator preserves the $\mathbb{Z}_{2}$ symmetry near the boundary. 
In contrast, at the cavity cluster-Ising superradiant QCP, the condensation of the $\mathbb{Z}_2^T$-charged $\mu$ operator is forbidden, thereby stabilizing a boundary condition with spontaneously broken $\mathbb{Z}_2$ symmetry and an associated twofold topological degeneracy. 
This distinction in boundary conditions accounts for the different boundary critical behaviors observed in Fig.~\ref{fig2}(b). 
As an independent diagnostic, we further examine the energy spectrum $\{\epsilon_n\}$ under OBC at the two transition points; see Fig.~\ref{fig3}(b) and Fig.~\ref{fig3}(c). 
A robust twofold degeneracy is observed throughout the energy spectrum at the cavity cluster-Ising superradiant QCP (red squares), whereas no such degeneracy is observed at the cavity Ising superradiant QCP (blue circles). 
The twofold degeneracy persists in the presence of symmetry-preserving disorder and perturbations, further confirming its topological origin (see Supplementary Note~2).

\vspace{.5cm}
\noindent\textbf{\large{}Generalization to the cavity Potts chain}

\noindent
Moreover, we show that the topological superradiant criticality uncovered here is not restricted to the $\mathbb{Z}_{2}$ Ising case but also occurs in $\mathbb{Z}_{n}$-symmetric cluster-qudit chains coupled to a cavity mode.
In this broader setting, the topological superradiant universality class emerges in the critical regime connecting the $\mathbb{Z}_n \times \mathbb{Z}_n$ cluster SPT phase and the superradiant phase with spontaneously broken $\mathbb{Z}_n$ parity symmetry.

We now test this general phenomenon numerically using the $\mathbb{Z}_3\times\mathbb{Z}_3$ cluster-Potts chain coupled to a single photonic mode as an illustrative example. 
On a bipartite chain with sublattices $A$ and $B$, we consider
\begin{equation}
  \label{E2}
  \begin{split}
  H_{\mathrm{CP}}^{\mathrm{cavity}} & = \omega_0 a^{\dagger}_{} a^{}_{} + \frac{g_3}{\sqrt{L}} \left( a^{}_{} \sum_i Z^{}_{B,i} + a^{\dagger}_{} \sum_i Z^{\dagger}_{B,i} \right) + H_{\mathrm{CP}} \,, \\
  H_{\mathrm{CP}} & = - J_{3} \sum_{i} Z_{B,i}^{\dagger} Z_{B,i+1}^{} - f_3 \sum_i \big( Z^{\dagger}_{B,i-1} X^{}_{A,i} Z^{}_{B,i} \\
  & \quad{} + Z^{}_{A,i} X^{}_{B,i}  Z^{\dagger}_{A,i+1} \big) + \mathrm{H.c.} \, ,
  \end{split}
\end{equation}
where
\begin{equation}
  X = \begin{pmatrix}
    0 & 1 & 0 \\
    0 & 0 & 1 \\
    1 & 0 & 0
  \end{pmatrix}
  \quad \text{and} \quad
  Z = \begin{pmatrix}
    1 & 0 & 0 \\
    0 & \omega & 0 \\
    0 & 0 & \omega^2 
  \end{pmatrix}
\end{equation}
with $\omega = \text{e}^{2\pi \text{i}/3}$ and $X Z = \omega Z X$. 
The Hamiltonian is invariant under a $\mathbb{Z}_{3}^{A}$ symmetry generated by $P_{A} = \prod_i X^{}_{A,i}$ and a joint $\mathbb{Z}_3$ symmetry generated by $\mathcal{P}_{3} = \omega^{a^\dagger_{} a^{}_{}} \prod_i X^{}_{B,i}$. 
At $g_3 = 0$, the photonic mode decouples from the qutrit degrees of freedom, and the Potts chain exhibits a topological continuous transition at $f_3/J_3 = 1.0$, separating a $\mathbb{Z}_3^B$ symmetry-breaking phase from a $\mathbb{Z}_3^A \times \mathbb{Z}_3^B$ SPT phase described by the $(1+1)$-dimensional Potts CFT with central charge $c = 4/5$~\cite{di1997conformal,ginsparg1988appliedconformalfieldtheory}. 
Starting from the SPT phase at fixed $f_3/J_3 = 1.2$ and gradually increasing the light-qutrit coupling $g_3$, the system undergoes a continuous superradiant transition from the $\mathbb{Z}_3$ cavity SPT to a superradiant phase that spontaneously breaks the joint $\mathbb{Z}_3$ symmetry, as confirmed by the photon occupation number vanishes continuously from one side to the another [Fig.~\ref{fig4}(a)], together with logorithmic scaling behavior of spatial qutrit entanglement entropy [Fig.~\ref{fig4}(b)] at the critical point identified by the crossing of the qutrit Binder cumulant [inset of Fig.~\ref{fig4}(a)]. 
Furthermore, this $\mathbb{Z}_3$ superradiant transition is topologically distinct from the conventional ones (see Supplementary Note~4 for additional numerical results), as evidenced by the slower power-law decay of the $\mathbb{Z}_3$ SPT string operator [Fig.~\ref{fig4}(c)] and the robust threefold topological degeneracy of the open boundary energy spectrum [Fig.~\ref{fig4}(d)].

\vspace{.5cm}
\noindent\textbf{\large{}Discussion and concluding remarks}

\noindent
In Supplementary Note~3, we further examine additional representative points along the phase boundary between the cavity SPT and the superradiant phase in the cavity cluster-Ising chain. 
These points exhibit similar topological and critical properties discussed in the main text, suggesting that the entire phase boundary belongs to the topological superradiant universality class. 

To summarize, we reveal a topological superradiant universality class unique to light-matter coupled systems. 
By numerically solving the cavity cluster-Ising chain, we map out its global ground-state phase diagram, which hosts a cavity SPT phase and a superradiant phase characterized by a macroscopic photon occupation. 
More importantly, we unambiguously show that global symmetry can enrich this phase boundary, giving rise to a novel topological superradiant universality class distinct from previously recognized superradiant transitions. 
This phenomenon is generic and also emerges in cavity-mediated $\mathbb{Z}_n$ cluster-qudit chains, as demonstrated by complementary lattice simulations. 

Several directions follow naturally from this work. It would be particularly interesting to investigate whether the topological superradiant universality class persists in other light-matter coupled systems, such as cavity-coupled electronic systems, and to explore its possible extensions to higher spatial dimensions. 
On the experimental side, the cluster spin model has recently been realized in ultracold-atom~\cite{Petiziol2021PRL,Sylvain2019Science,samajdar2026threebodyinteractionsrydberglattices} and superconducting qubit quantum platforms~\cite{Zhang2022Nature,TanCP2026}, where cavity-mediated couplings can also be conveniently engineered and controlled within the same settings.

\vspace{.5cm}
\noindent\textbf{\large{}Methods}{\large\par}

\noindent\textbf{Details of the DMRG calculations for the cavity-mediated spin chains}

\noindent 
We employ the finite-size density-matrix renormalization group (DMRG)~\cite{PhysRevLett.69.2863,PhysRevB.48.10345} to obtain the ground-state wave function, correlation functions, string order parameter, entanglement entropy, and low-lying excitation spectrum of the system. 
DMRG is a variational numerical method particularly suited to one-dimensional and quasi-one-dimensional quantum many-body systems. 
It represents the many-body wave function as a matrix product state (MPS)~\cite{PhysRevB.73.094423,SCHOLLWOCK201196,ORUS2014117}, 
\begin{equation}
    |\Psi\rangle = \sum_{\sigma_1,\dots,\sigma_N} A^{\sigma_1} A^{\sigma_2}\cdots A^{\sigma_N} |\sigma_1\sigma_2\cdots\sigma_N\rangle \, ,
\label{eq:mps}
\end{equation}
where $A^{\sigma_i}$ are site tensors with bond dimension $\chi$. 
The DMRG algorithm sweeps through the entire lattice and locally optimizes these tensors to minimize the energy
\begin{equation}
    E=\frac{\langle\Psi| H|\Psi\rangle}{\langle\Psi | \Psi\rangle} \, .
\end{equation}
At each bond, we retain the largest Schmidt coefficients, or singular values $\lambda_{\alpha}$, up to a prescribed bond dimension $\chi$ or a truncation threshold. 
The discarded weight
\begin{equation}
  \epsilon_\text{cut} = 1-\sum_{\alpha=1}^{\chi}\lambda_\alpha^2 \, ,
\end{equation}
provides a controlled measure of the truncation error. 
For gapped one-dimensional systems, the entanglement entropy obeys the entanglement area law and saturates with system size, so a moderate bond dimension is often sufficient. 
For critical systems, the entanglement entropy grows logarithmically with the system size, requiring a larger $\chi$. 
The DMRG sweeps are considered converged when the change in the ground-state energy between two successive sweeps satisfies $\left|E_{k}-E_{k-1}\right|<\delta_\text{conv}$, where $k$ denotes the sweep index, and $\delta_{\mathrm{conv}}$ the energy convergence tolerance.

To treat the cavity-mediated spin chain within the MPS framework, we map the single photonic mode to an auxiliary bosonic lattice site. 
Since the photon Fock space is infinite, we truncate it at a maximum occupation number $N_\mathrm{cutoff}$, so that the local Hilbert space of the photon site is spanned by $|0\rangle,|1\rangle,\dots,|N_\mathrm{cutoff}-1\rangle$. 
The total system is then represented as an MPS with one bosonic site and $L$ spin-$1/2$ sites. 
For the cavity Potts chain, the spin sites are replaced by qutrit sites with local dimension $d = 3$. 
In practice, we place the photon site at one end of the chain. 
Taking the Ising model as an example, the site ordering is ``$\mathrm{photon},\sigma_1,\sigma_2,\dots,\sigma_L$'', or the reverse. 
The Hamiltonian can be written in the form 
\begin{equation}
    H=H_{\mathrm{ph}} \otimes I_{\mathrm{spin}}+I_{\mathrm{ph}} \otimes H_{\mathrm{spin}}+\frac{g}{\sqrt{L}}\left(a^{}_{} + a_{}^{\dagger}\right) \otimes \sum_{i} \sigma_{i}^{x} \, ,
\end{equation}
with $H_{\mathrm{ph}} = \omega_0 a_{}^\dagger a_{}^{}$. 
This structure is convenient for MPO construction: the global light-matter coupling factorizes into a single photon operator $\left(a+a^\dagger\right)$ acting on the auxiliary bosonic site and the spin operator $\sum_i\sigma_i$ acting on the matter chain.
The latter can be represented by a standard MPO with a small bond dimension along the spin chain, while the photon operator acts only on the bosonic site.

After convergence, the ground-state MPS $|\psi_0\rangle$ is used to compute physical observables by contracting the corresponding tensor network. 
To obtain the entanglement entropy, we divide the system into subsystems $\mathcal{L}$ and $\mathcal{R}$ and perform a Schmidt decomposition,
\begin{equation}
    \left|\psi_{0}\right\rangle = \sum_{\alpha} \lambda_{\alpha}\left|\alpha_{\mathcal{L}}\right\rangle\left|\alpha_{\mathcal{R}}\right\rangle \, ,
\end{equation}
and the von Neumann entropy is
\begin{equation}
    S_{\mathcal{L}} = -\sum_{\alpha} \lambda_{\alpha}^{2} \ln \lambda_{\alpha}^{2} \, .
\end{equation}
In the MPS representation, the $\lambda_\alpha$ are directly related to the singular values retained at the corresponding bond.

Low-lying excited states are computed using orthogonal DMRG, or the penalty method. 
After obtaining the lower-lying states, we minimize the energy in their orthogonal complement by adding a penalty term to the Hamiltonian. 
To obtain the $n$-th excited state $|\psi_n\rangle$, with $|\psi_0\rangle$ the ground state, after the states $|\psi_0\rangle,\dots,|\psi_{n-1}\rangle$ have been obtained, we use 
\begin{equation}
    H^{(n)}=H + w\sum_{i=0}^{n-1} |\psi_i\rangle\langle\psi_i| \, .
\end{equation}
The second term is the projector onto the subspace spanned by the previously obtained states; 
it shifts these states upward in energy by $w$. 
For $w$ sufficiently large compared with the energy differences between the target state and the lower-lying states, the ground state of $H^{(n)}$ is $|\psi_n\rangle$. 
A standard DMRG ground-state search for $H^{(n)}$ then yields the next energy level.

In our DMRG calculations, the bond dimension $\chi$ is taken to be sufficiently large to ensure convergence of the ground-state and low-lying excited-state properties. 
At each singular-value-decomposition truncation, we keep the discarded weight below $\epsilon_\text{cut} = 10^{-10}$, stop the DMRG sweeps when the energy difference between successive sweeps falls below $\delta_{\mathrm{conv}} = 10^{-8}$, and use orthogonal DMRG with a penalty term with $w=50$ for the low-lying excited states.

\vspace{.5cm}
\noindent\textbf{Topological nontrivial QCPs.}

\noindent 
Here we briefly review the physics of topological nontrivial criticality (also called symmetry-enriched quantum criticality in the literature~\cite{Verresen2021PRX}) by considering the quantum phase transitions in the transverse-field Ising and cluster-Ising chains, described respectively by the Hamiltonian
\begin{equation}
  \begin{split}
  H_\text{TFIM} & = -J \sum_{i} \sigma_{i}^{x} \sigma_{i+1}^{x} - h \sum_{i} \sigma_{i}^{z} \, , \\
  H_\text{CI} & = - J \sum_{i} \sigma^{x}_{i} \sigma^{x}_{i+1} - f \sum_{i} \sigma^{x}_{i} \sigma^{z}_{i+1} \sigma^{x}_{i+2} \, .
  \end{split}
\end{equation}
Both models possess a global spin-flip $\mathbb{Z}_2$ symmetry and a spinless time-reversal symmetry $\mathbb{Z}_2^T$ generated by complex conjugation $K$.
Their critical points ($h/J = 1$ and $f/J = 1$) are described by the same $(1+1)$-dimensional Ising CFT, which features the local spin field $\sigma$ and the nonlocal disorder field $\mu$, both with the scaling dimension $\Delta_\sigma = \Delta_\mu = 1/8$.
At the microscopic lattice level, $\sigma_i$ is represented by the local magnetic order parameter $\sigma_i^x$, while $\mu_i$ corresponds to its Kramers--Wannier dual string order parameter of the symmetric phase.
For the transverse-field Ising chain, it is represented by a semi-infinite string
\begin{equation}
  \mu_i \propto \prod_{m<i} \sigma_m^z \, .
  \label{eq:ising-disorder-field}
\end{equation}
For the cluster-Ising chain, the corresponding disorder field has a nontrivial endpoint decoration,
\begin{equation}
  \widetilde \mu_i \propto \left( \prod_{m<i-1} \sigma_m^z \right) \sigma_{i-1}^y \sigma_i^x \, .
  \label{eq:cluster-ising-disorder-field}
\end{equation}
Under the spinless time-reversal symmetry, these operators carry opposite charges,
\begin{equation}
  K \mu_i K = \mu_i \, , \qquad K \widetilde\mu_i K = -\widetilde\mu_i \, .
  \label{eq:ising-flux-charges}
\end{equation}
Therefore, the $\mathbb{Z}_2^T$ charge of the disorder field provides a \textit{discrete invariant} that distinguishes the conventional Ising and topological Ising QCPs despite their identical local bulk scaling dimensions.
This distinction is also reflected in their boundary physics.
Under their natural open-boundary terminations, the transverse-field Ising chain realizes the free conformal boundary condition, whereas the cluster-Ising chain realizes the spontaneously fixed boundary condition and hosts a twofold edge degeneracy~\cite{Yu2022PRL,Parker2018PRB}.
The two critical points can therefore be distinguished through their nonlocal string correlations, boundary critical behavior, and open-boundary energy-spectrum degeneracies~\cite{Yu2024PRL,Zhang2024PRA}.

\vspace{.5cm}
\noindent\textbf{Observables for the cavity Ising chains}

\noindent 
We first define the observables used to characterize the cavity transverse-field Ising and cavity cluster-Ising chains.
The photon occupation and its intensive density are
\begin{equation}
  N_\text{ph} = \langle a^\dagger a^{} \rangle \, , \qquad n_\text{ph} = \langle a^\dagger a^{} \rangle / L \, .
  \label{eq:photon-number}
\end{equation}
In a superradiant phase, $N_\text{ph}$ is macroscopic and hence $n_\text{ph}$ remains finite in the thermodynamic limit. 
We also identify the continuous superradiant transition from the peak of $- \partial N_\text{ph}/\partial f$.

For a lattice operator $\phi_i$, we calculate the connected bulk-bulk and boundary-bulk correlation functions
\begin{align}
  C_{\phi}^\text{bulk}(r) & = \langle \phi_{i} \phi_{i+r} \rangle - \langle \phi_{i} \rangle \langle \phi_{i+r} \rangle \, ,
  \label{eq:bulk-correlation-general} \\
  C_{\phi}^\text{bdy}(r) & = \langle \phi_{1} \phi_{1+r} \rangle - \langle \phi_{1} \rangle \langle \phi_{1+r} \rangle \, .
  \label{eq:boundary-correlation-general}
\end{align}
The bulk-bulk correlations are evaluated under PBC, whereas the boundary-bulk correlations are evaluated under OBC.
For the cluster-Ising chain, however, since the boundary operator $\sigma_{1}^{x}$ commutes with the OBC Hamiltonian, we evaluate the spin-spin boundary-bulk correlation between sites $2$ and $2+r$~\cite{Yu2022PRL};
for the transverse-field Ising chain, we use the definition in Eq.~\eqref{eq:boundary-correlation-general} directly.
At a scale-invariant QCP, their leading long-distance behaviors should obey power-law forms
\begin{equation}
  C_{\phi}^\text{bulk}(r) \sim r^{-2\Delta_{\phi}^\text{bulk}} \, , \qquad C_{\phi}^\text{bdy}(r) \sim r^{-(\Delta_{\phi}^\text{bdy} + \Delta_{\phi}^\text{bulk})} \, .
  \label{eq:correlation-scaling}
\end{equation}
In practice, we choose $r=L/2$ throughout the work. 
For the spin-spin correlation, $\phi_{i} = \sigma_{i}^{x}$, and we denote the resulting correlations by $C_\text{FM}^\text{bulk/bdy}(r)$, whereas for the energy-energy correlation, we use the lattice energy-density operator, $\sigma_{i}^{x} \sigma_{i+1}^{x}$, and denote the corresponding connected correlations by $C_\text{EN}^\text{bulk/bdy}(r)$.

Nonlocal string operators probe the symmetry-resolved information that is invisible to local bulk correlations.
For the cavity Ising chains, the two string order parameters considered in the main text are
\begin{align}
  O_\text{PM}(r) & = \left\langle \text{e}^{\text{i} \pi a^\dagger a^{}} \left( \prod_{k=i}^{i+r-1} \sigma_{k}^{z} \right) \right\rangle \, , \label{eq:opm} \\
  O_\text{SPT}(r) & = \left\langle \text{e}^{\text{i} \pi a^\dagger a^{}} \sigma_{i}^{x} \sigma_{i+1}^{y} \left( \prod_{k=i+2}^{i+r-1} \sigma_{k}^{z} \right) \sigma_{i+r}^{y} \sigma_{i+r+1}^{x} \right\rangle \, . \label{eq:ospt}
\end{align}
The photon-parity factor combines with the Pauli string because the $\mathbb{Z}_{2}$ part of the protecting symmetry of the coupled system is the joint action $\mathcal{P} = \text{e}^{\text{i} \pi a^\dagger a^{}} \left( \prod_{i} \sigma_{i}^{z} \right)$.
In the decoupled cavity spin chains, $O_\text{PM}$ and $O_\text{SPT}$ are the two-point functions of the disorder fields appropriate to the trivial and cluster realizations, respectively.

To study the entanglement properties of the ground state $|\Psi\rangle$ of the light-matter system, we consider two complementary bipartitions that probe the spatial and photon-spin entanglement, respectively.
To characterize the spatial entanglement, we partition the system such that subsystem $\mathcal{L}$ consists of the cavity photon mode together with the first half of the spin chain, while its complement $\mathcal{R}$ contains the remaining spins.
Defining the reduced density matrix $\rho_{\mathcal{L}} = \text{Tr}_{\mathcal{R}} \left( |\Psi\rangle \langle\Psi| \right)$, the corresponding von Neumann entanglement entropy is
\begin{equation}
  S_\text{vN} = - \text{Tr} \left( \rho_{\mathcal{L}} \ln \rho_{\mathcal{L}} \right) \, .
  \label{eq:spatial-entanglement}
\end{equation}
For a critical chain under PBC, we fit the finite-size scaling of the half-chain entanglement entropy to~\cite{Calabrese_2009}
\begin{equation}
  S_\text{vN}(L) = \frac{c_\text{eff}}{3} \ln L + s_{0} \, ,
\end{equation}
where $c_\text{eff}$ is a coefficient that characterizes the logarithmic growth rate of the entanglement entropy with system size and $s_{0}$ is a nonuniversal constant.
To characterize the entanglement between the light and matter degrees of freedom, we instead bipartition the system into the cavity photon mode and the entire spin chain.
Defining $\rho_\text{ph} = \text{Tr}_\text{spin} \left( |\Psi\rangle \langle\Psi| \right)$, the photon-spin entanglement entropy is given by
\begin{equation}
  S_\text{PS} = - \text{Tr} \left( \rho_\text{ph} \ln \rho_
  \text{ph} \right) \, .
  \label{eq:photon-spin-entanglement}
\end{equation}
Since $|\Psi\rangle$ is a pure state, $S_\text{PS}$ is equivalently obtained from the reduced density matrix of the entire spin chain.

\vspace{.5cm}
\noindent\textbf{Corresponding observables for the cavity Potts chain}

\noindent
We now introduce the corresponding symmetry and observables for the cavity Potts chain, following the $\mathbb{Z}_{3}^{A} \times \mathbb{Z}_{3}^{B}$ construction of Ref.~\cite{Verresen2021PRX}.
The whole chain is divided into unit cells labeled by $i$, each containing two qutrits, respectively, denoted by $A$ and $B$.
In the absence of the cavity, the two sublattice symmetries are independently generated by $P_{A} = \prod_i X_{A,i}$ and $P_{B} = \prod_i X_{B,i}$.
The light-qutrit coupling preserves $P_{A}$, while $P_{B}$ is promoted to the joint action $\mathcal{P}_{3} = \omega^{a^\dagger a^{}} P_{B}$.
Accordingly, $Z_{B,i}$ serves as the local order parameter for the symmetry that is spontaneously broken in the superradiant phase, while the $A$ sublattice carries the nontrivial endpoint charge associated with the SPT symmetry flux.
Here, $L$ denotes the number of unit cells, so that the matter sector contains $2L$ qutrits.
The photon observables $N_\text{ph}$ and $n_\text{ph}$, as well as the two entanglement entropies, are defined in the same way as in Eqs.~\eqref{eq:photon-number}, \eqref{eq:spatial-entanglement}, and~\eqref{eq:photon-spin-entanglement}, with the spatial subsystem $\mathcal{L}$ now consisting of the photon mode together with the first $L/2$ complete unit cells.

We further define the corresponding magnetization and its Binder cumulant~\cite{sandvik2010AIP}
\begin{equation}
  M_{B}^{} = \frac{1}{L} \sum_{i=1}^{L} Z_{B,i}^{} \, , \quad U_{4}^\text{Potts} = 2 - \frac{\left\langle M_{B}^\dagger M_{B}^{} M_{B}^\dagger M_{B}^{} \right\rangle}{\left\langle M_{B}^\dagger M_{B}\right\rangle^{2}} \, .
  \label{eq:potts-binder}
\end{equation}

The distinction between the conventional and topological nontrivial Potts critical points is most transparent from the associated symmetry flux.
A left semi-infinite flux terminating near unit cell $i$ has the form~\cite{Verresen2021PRX}
\begin{equation}
  \mathcal{S}_{i}^{B} = \cdots X_{B,i-2}X_{B,i-1} \mathcal{O}_{i}^{B} \, .
  \label{eq:potts-symmetry-flux}
\end{equation}
The endpoint operator is trivial, $\mathcal{O}_{i}^{B} = I$, in the conventional realization;
while in the symmetry-enriched realization, it is decorated by an operator on the sublattice $A$, $\mathcal{O}_{i}^{B} = Z_{A,i}^\dagger$, which carries a nontrivial charge under $P_A$.
The corresponding string order parameters (i.e., two-point functions of the symmetry fluxes) in the cavity Potts system are
\begin{align}
  O_\text{PM}^\text{Potts}(r) & = \left\langle \omega^{a^\dagger a^{}} \left( \prod_{k=i}^{i+r-1} X_{B,k} \right) \right\rangle \, , \label{eq:potts-pm-string} \\
  O_\text{SPT}^\text{Potts}(r) & = \left\langle \omega^{a^\dagger a^{}} Z_{A,i} \left( \prod_{k=i}^{i+r-1} X_{B,k} \right) Z_{A,i+r}^\dagger \right\rangle \, . \label{eq:potts-spt-string}
\end{align}
At the topological Potts superradiant QCP, the leading algebraic decay of $O_\text{SPT}^\text{Potts}$ and the threefold degeneracy of ground-state manifold provide the direct counterparts of the string and spectrum diagnostics used for the cavity cluster-Ising chain.

\vspace*{.6cm}
\noindent\textbf{\large{}Data availability}
All the numerical codes and data used in the study are available from the corresponding author (X.-J. Yu) upon reasonable request.

\vspace{.5cm}
\noindent\textbf{Acknowledgement}
X.-J. Yu was supported by the National Natural Science Foundation of China (Grant No.12405034) and a start-up grant from Eastern Institute of Technology, Ningbo. The work of S.Y. is supported by China Postdoctoral Science Foundation (Certificate Number: 2024M752760). Tensor-network simulations were carried out with the ITENSOR \verb|C++| package~\cite{itensor}.

\vspace{.3cm}
\noindent\textbf{Author contributions} X.-J. Y. conceived the research. Y.L., S.Y., M.C. carried out model
studies. X.-J.Y. Y. L. and S.Y. wrote the paper, with inputs from all the other authors (M.C.).

\vspace{.3cm}
\noindent\textbf{Competing interests}  All authors declare no competing interests.

\bibliography{main}

\end{document}